\documentclass[12pt,preprint]{aastex}

\newcommand{\be}{\begin{equation}}
\newcommand{\etal}{{\em et al. }}
\newcommand{\ltorder}{\hbox{ \rlap{\raise 0.425ex\hbox{$<$}}\lower
0.65ex\hbox{$\sim$} }}
\newcommand{\gtorder}{\hbox{ \rlap{\raise 0.425ex\hbox{$>$}}\lower
0.65ex\hbox{$\sim$} }}
\newcommand{\mbh}{M_{BH}}
\shorttitle{Surface brightness cusps in globular clusters}
\shortauthors{Vesperini and Trenti}
\begin{document}

\title{Widespread presence of shallow cusps in the surface-brightness profile of globular clusters}
\author{Enrico Vesperini}
\affil{Department of Physics, Drexel University, Philadelphia, PA 19104, USA}
\email{vesperin@physics.drexel.edu}
\and
\author{Michele Trenti}
\affil{University of Colorado, CASA, Dept.  of Astrophysical \& 
Planetary Sciences, 389-UCB, Boulder, CO 80309, USA} 
\email{trenti@colorado.edu}
\begin{abstract}

  Surface brightness profiles of
  globular clusters with shallow central cusps ($\Sigma\sim R^{\nu}$
  with $-0.3\ltorder \nu\ltorder -0.05$) have
  been associated by several recent studies with the presence of a
  central intermediate mass black hole (IMBH). Such shallow slopes
  are observed in several globular clusters thanks to the
  high angular resolution of Hubble
  Space Telescope imaging. In this Letter we evaluate whether shallow
  cusps are a unique signature of a central IMBH by analyzing a sample
  of direct N-body simulations of star clusters with and without a
  central IMBH. We ``observe'' the simulations as if they were HST
  images. Shallow cusps are common in our simulation sample:
  star clusters without an IMBH have $\nu\gtorder -0.3$
  in the pre-core-collapse and core-collapse phases.
  Post-core-collapse clusters without an IMBH transition to steeper
  cusps,  $-0.7\ltorder \nu \ltorder -0.4$, only if the primordial binary
  fraction is very small, $f_{bin}< 3\%$, and if there are few
  stellar-mass black holes remaining. Otherwise $\nu$ values overlap the
  range usually ascribed to the presence of an IMBH
  throughout the entire duration of the simulations. In addition,
  measuring $\nu$ is intrinsically prone to significant uncertainty,
  therefore typical measurement errors may lead to $\nu\geq -0.3$ even
  when $\langle\nu\rangle\lesssim -0.4$. Overall our
  analysis shows  that   a shallow cusp is {\it not} an
  unequivocal signature of a central IMBH and casts
  serious doubts on the usefulness of measuring $\nu$ in the context
  of the hunt for IMBHs in globular clusters.
\end{abstract}

\section{Introduction}
\label{sec:introd}
Several observational studies have provided strong evidence for the
presence of supermassive black holes with masses $M_{BH}$ in the range
$10^6~\ltorder~M_{BH}/M_{\odot}~\ltorder~10^9$ at the centers of elliptical and
spiral galaxies.  The masses of these central black holes 
correlate with the 
host-galaxy bulge mass and velocity dispersion (see e.g. G\"ultekin et
al. 2009).  If these correlations  
are extrapolated to smaller stellar systems, globular clusters should
harbor intermediate-mass black holes (hereafter IMBHs) with masses in
the range $10^2-10^4~M_{\odot}$.

Despite significant observational efforts, conclusive evidence of the
presence of IMBHs in clusters is still lacking. Observational studies
of Galactic clusters have so far provided only upper limits on the
masses of  IMBHs (see van der Marel \& Anderson 2010 and references
therein). The difficulty of obtaining an unambiguous detection is
illustrated by the Omega Cen case. Based on integrated light
spectroscopy from the ground and an HST-based
surface brightness profile , Noyola \etal (2008) claimed a
$4.0^{+0.75}_{-1.0}~\times~10^4~M_{\sun}$ IMBH. 
However, proper-motion-velocity-dispersion data obtained with HST (van
del Marel \& Anderson,  
Anderson \& van der Marel 2010) show that $\mbh~\ltorder~1.8\times~10^4~M_{\odot}$ (at $3\sigma$ confidence) in contrast with the
Noyola \etal result. Outside the Milky Way, the presence of a central
IMBH has been 
suggested for G1, a massive cluster in M31. Here radio, X-ray
observations and dynamical studies all suggest the presence of an IMBH
with $\mbh\sim 2 \times 10^4~M_{\odot}$ (Gebhardt \etal 2005,
Ulvestad \etal 2007), but because of the limited spatial resolution
of the observations (G1 is at about 800 Kpc, compared to the $\lesssim 10$
Kpc for many Galactic clusters), it is hard to probe the central
region with proper motions or star-by-star radial velocities to
evaluate the robustness of the G1 IMBH detection. 

Given the difficulty of achieving a definitive detection of an IMBH,
several studies have focused on possible signatures for its presence,
both to construct circumstantial evidence based on a coincidence of
indicators, as well as to identify a sample of promising candidates
for detailed kinematic follow-up. In young clusters, X-ray (ULX)
emission due to gas accretion is expected (see e.g. Miller \& Colbert
2004). Dynamic  
signatures are related to the formation of a Bahcall-Wolf density
cusp (Bahcall \& Wolf 1976) within the sphere of influence of the BH. The
three-dimensional 
density cusp induces a shallow cusp ( $\Sigma~\sim~R^{\nu}$, with
$ -0.3~\ltorder~\nu~\ltorder~-0.05$) in the projected surface brightness and
velocity dispersion (Baumgardt \etal 2005, Trenti \etal 2007a,
Miocchi 2007, Umbreit \etal 2009). In addition, these
numerical studies  showed that IMBHs can act as a
very efficient energy source in clusters because of enhanced
three-body interactions within the BH sphere of influence, supporting
large  
cores and preventing core-collapse. Finally, Gill \etal (2008) have
shown that the presence of an IMBH in a globular cluster can quench the
process of mass segregation and introduced a new IMBH indicator based
on the radial profile of the mean star mass within a cluster. This
indicator has been successfully applied to two globular clusters
NGC2298 (Pasquato \etal 2009) and M10 (Beccari \etal 2010), ruling
out a $M>300~M_{\odot}$ IMBH (3-$\sigma$ confidence) in the first case
and providing weaker limits in the second.

An important property for a reliable IMBH fingerprint is its
uniqueness. For example, while systems in which a small
core-to-half-mass radius ratio is observed are unlikely to host a
IMBH, the 
converse is not necessarily true. Other dynamic processes, primarily
two-body interactions of dark remnants with main sequence stars,
provide efficient heating on the visible population of stars and may mask
the development of core collapse (Trenti \etal 2010).  Motivated by
the recent attention given to the presence  
of shallow central cusps as fingerprints of IMBH in a number of
observed Galactic and Magellanic Clouds clusters (see e.g. Noyola \&
Gebhardt 2006, 2007, Lanzoni \etal 2007), we discuss in this Letter
the likelihood of observing shallow cusps in systems without an IMBH.
For this, we take a set of direct N-body simulations of star
clusters and analyze their surface-brightness profiles as we
were "observing" them with HST (see Trenti et al. 2010).

The outline of this Letter is the following. Section~\ref{sec:initc}
summarizes the setup of our N-body simulations and defines the
analysis tools we use. Section~\ref{sec:res} presents the results on
the time evolution of the central slope in the surface brightness
profiles. In Section~\ref{sec:disc} we conclude with a discussion of
our results in the context of IMBH searches.

\section{Method and Initial Conditions}
\label{sec:initc}
\subsection{N-body simulations}
The simulations considered here are a subset of those presented in
Trenti \etal (2010). We
summarize here the main ingredients of our numerical framework.
Further details are provided in Trenti et al. (2010).

We follow the evolution of star clusters using the direct summation
code NBODY6 (Aarseth 2003) run with its GPU extension on the
NCSA Lincoln Cluster. The code models the effects of internal
relaxation with an accurate treatment of the multiple interactions
between stars obtained thanks to special regularization techniques.

The initial structure of the clusters is that of
tidally limited King (1966) models. In the simulations presented here,
the initial star masses are drawn either from a Miller \& Scalo (1979)
or a Salpeter (1955) initial mass function. Our analysis is focused
on the late evolutionary stages of clusters as driven by two-body
relaxation; before starting the run, we performed an
instantaneous step of stellar evolution to evolve the mass function to
a $0.8~\mathrm{M_{\sun}}$ turnoff using the Hurley et al. (2000)
evolutionary tracks. The evolution of the clusters has been followed
until either $t=8000$ (with the time expressed in N-body units,
equivalent to the cluster dynamical time) was reached (corresponding to several initial relaxation times
--- $t> 16~t_{rh}(0)$) or at least
80\% of the initial mass was lost due to evaporation of stars.

In order to explore the role of primordial hard binaries, our
sample of simulations includes systems without primordial binaries as
well as systems with an initial fraction of binaries
$f_{bin}=N_b/N$ (where $N_b$ is the number of binary stars) ranging from 0 to 0.1. The
binaries' binding energy distribution function is flat in log scale
between $\epsilon_{min}$ and $133\epsilon_{min}$ where
$\epsilon_{min}=\langle m(0)\rangle \sigma_c(0)^2$ where $\langle
m(0)\rangle$ is the initial mean stellar mass and $\sigma_c(0)$ is the
initial central velocity dispersion (see Trenti et al. 2007b, 2010).

In all the simulations we assumed that all the dark remnants
produced are retained in the cluster; the comparison of the results of
systems with a Miller-Scalo and a Salpeter initial mass function
(which have a similar fraction of white dwarfs and neutron stars but produce a significantly different fraction of stellar-mass 
black holes: $\sim 1.5\times 10^{-4}$ for Miller-Scalo
and $\sim 1.4\times 10^{-3}$ for Salpeter) allows us to explore the
dependence of our results on the fraction of massive dark remnants.
One simulation includes an IMBH with mass 1\% of the total cluster
mass.

The initial properties of the simulations considered here are
summarized in Table 1, along with the identification used below to
refer to them.

\subsection{Data Analysis}
In analysing the structural properties of our simulated clusters, we
follow a procedure aimed at reproducing as closely as possible
the steps performed in the analysis of observational data. We 
construct a synthetic observation by first ignoring all particles
except main sequence stars. As our simulations only have an
instantaneous step of stellar evolution, we do not consider stars
along the giant branch. This choice is appropriate to create a surface
brightness profile for main sequence  stars. Focusing on main sequence
stars leads to a 
significant reduction in the surface brightness profile fluctuations
arising from the small number of luminous giants that would otherwise
dominate the light profile. The mass difference between the upper main-sequence
stars we used and the giants should have a negligible effect on the surface-brightness profile.

We then select a random direction, project the main-sequence stars'
positions and create a 2-dimensional map. 
A circular surface brightness profile, $\Sigma(R)$, is constructed using
$2Int[\sqrt{N_{MS}}]$ annuli (where $N_{MS}$ is the number of main
sequence stars) each containing $\sim \sqrt{N_{MS}}/2$ sources. 
The center of the system is determined by using the Casertano \& Hut (1985)
density-center method. 
We then determine the King concentration
parameter $c=\log_{10}{(r_t/r_c)}$ by fitting the
surface brightness profile with a single-mass King (1966) model
(see Trenti et al. 2010 for further details on the construction of the
surface brightness profile and the fitting procedure and for the time
evolution of $c$).   

This Letter is focused on the evolution of the inner part of the
surface brightness profile. Specifically, we assume that in the
inner parts $\Sigma(R) \propto R^{\nu}$ and estimate $\nu$ by a fit
of $\Sigma(R)$ between the innermost $\Sigma(R)$ point (typically at
0.05-0.08 
$r_c$) and $r_c/3$ (or using at least the three innermost points)
where $r_c$ is the King core radius for the best-fit King
model. Previous simulations of clusters with an IMBH (Baumgardt et
al. 2005) show that shallow cusps extend to about $r_c/3$.  
In order to estimate the range of values of $\nu$
obtained with this procedure when a cluster structure is that of a
King model, we created 100 different random realizations of a
King model with $W_0=7$ and 32K particles. After calculating the
surface brightness profile for each system using the same number of
annuli used in the data analysis of our simulations, we 
calculated $\nu$ for each random realization using data points within
$r_c/3$. The mean value of $\nu$ obtained is $\sim -0.05$ with a
dispersion of $\sim 0.07$. Note that a small but non-zero value of
$\nu$ for a King model is to be expected. This can be easily verified also by adopting the King (1962) analytical expression 
$\Sigma=\Sigma_o/(1+(R/R_c)^2)$  to describe the inner surface density
profile of a King  model: fitting this analytical profile with a
power-law between $\sim 0.05r_c$  and $r_c/3$ yields $\nu\sim -0.05$. 
Smaller values of $\nu$  might be obtained by restricting the radial
range to regions closer to the center, but such an approach may be
complicated by 
fluctuations in the surface    
brightness profile of a cluster's innermost regions and, for actual data,
possibly by the lack of adequate resolution.  

Despite the different procedure adopted to measure the inner slope, 
our results are consistent within 1-sigma with the analysis of
artificial images of King models discussed in Noyola \& Gebhardt
(2006; see e.g. their results for $N=50000$ in  the 'subtracted' and '10\%
subtracted' panels of their Fig.4).

\section{Results}
\label{sec:res}
\begin{figure}
\begin{center}
\includegraphics[width=2.8in]{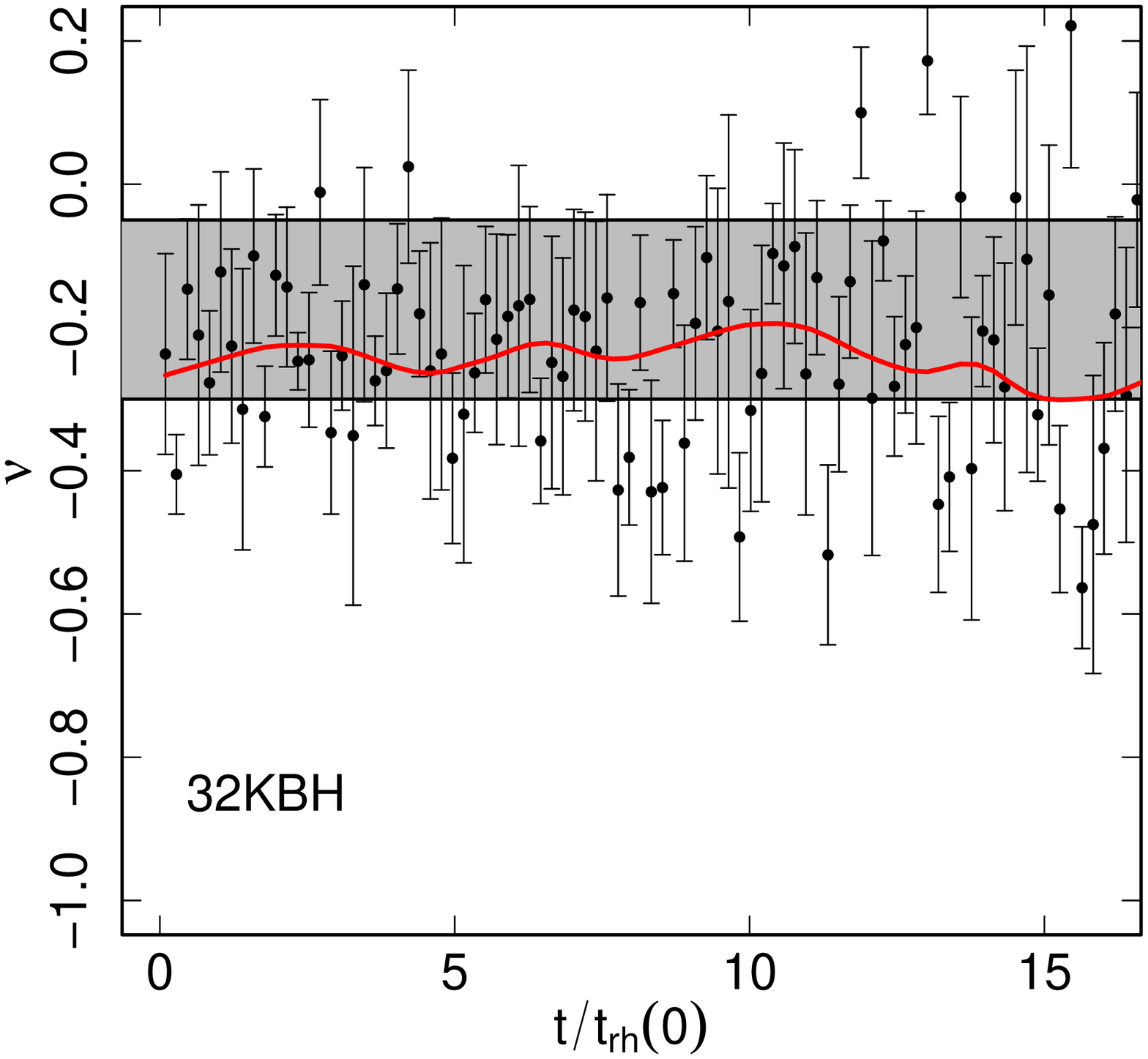}
\includegraphics[width=2.8in]{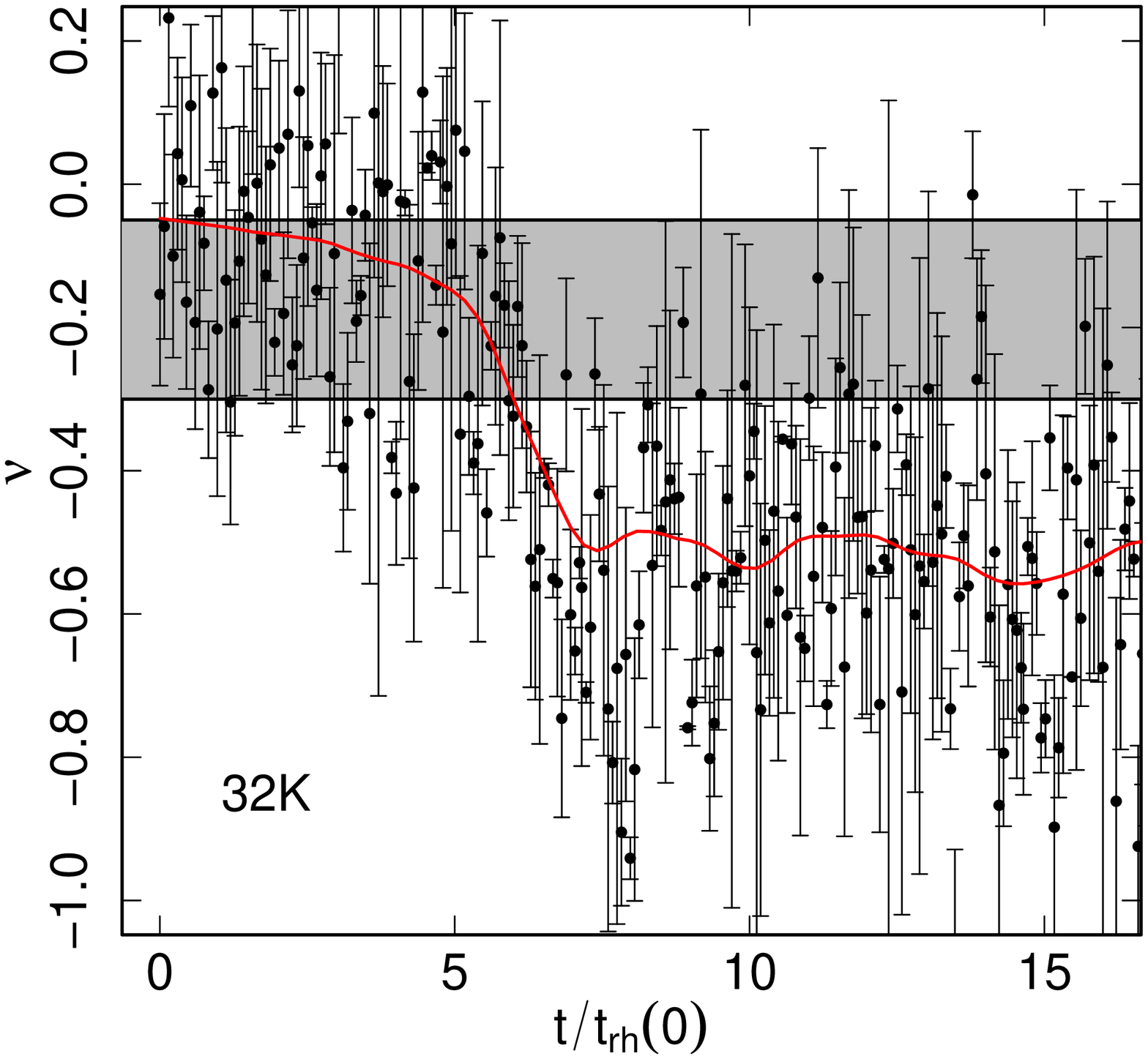}
\includegraphics[width=2.8in]{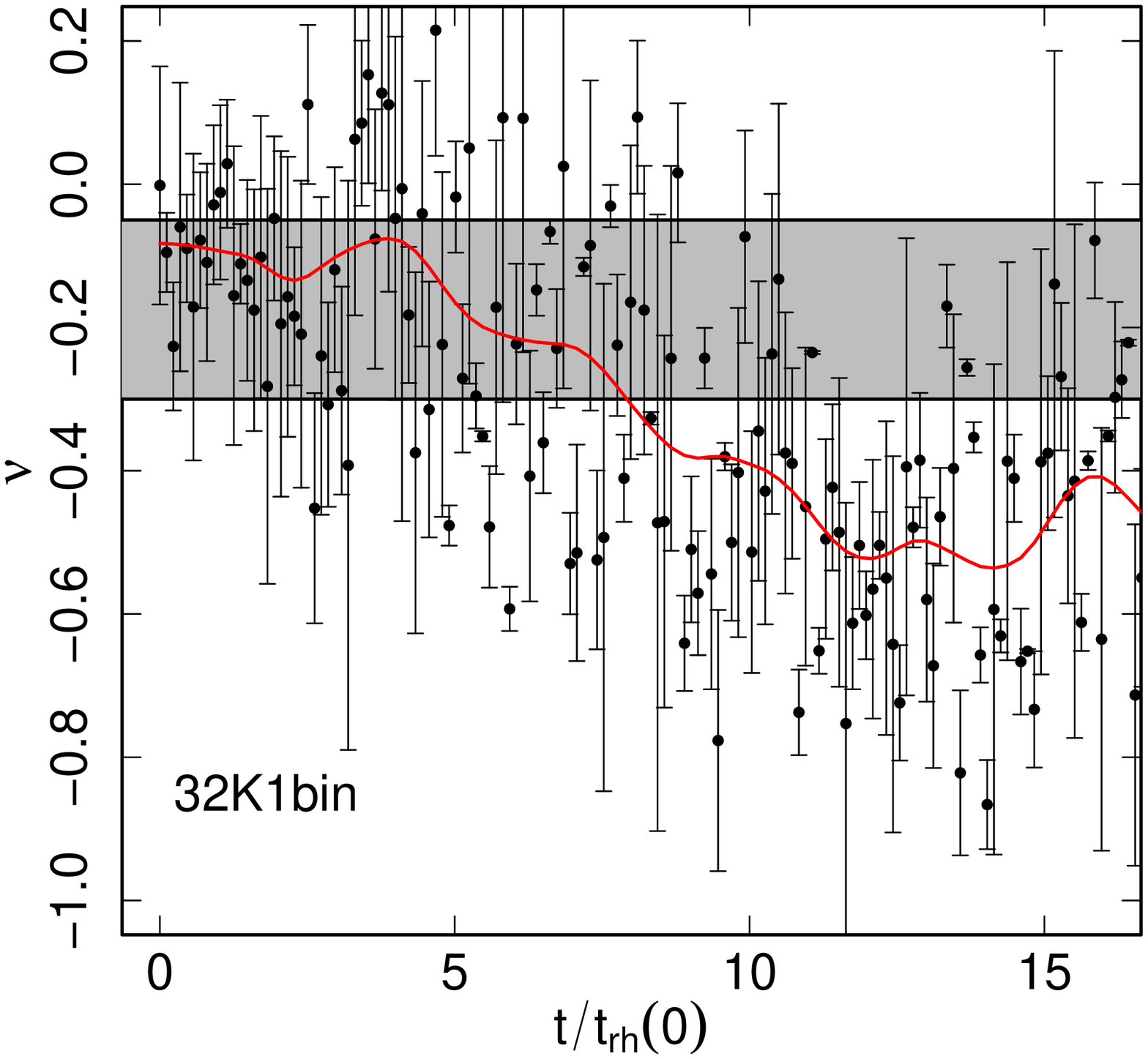}
\includegraphics[width=2.8in]{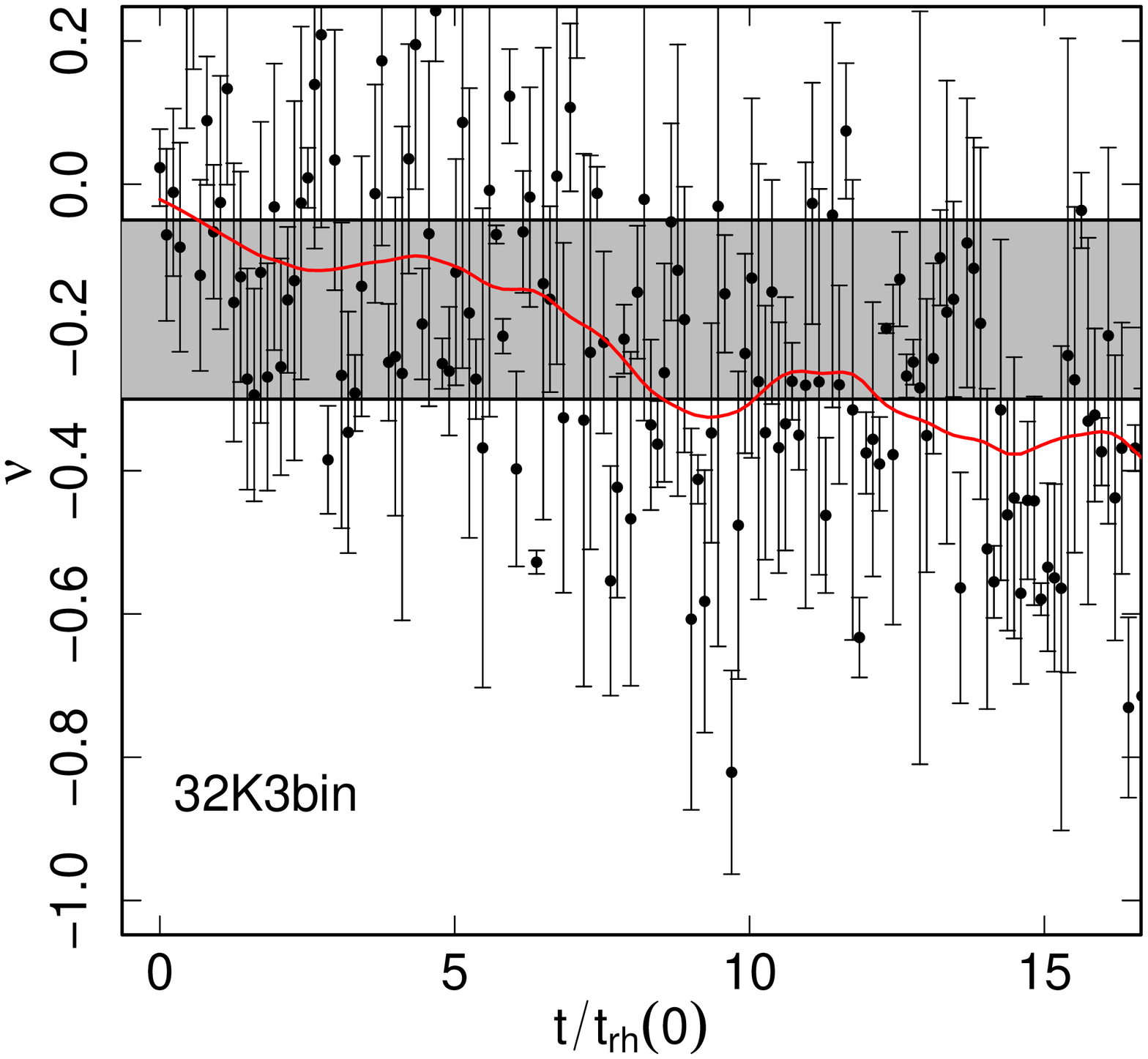}
\includegraphics[width=2.8in]{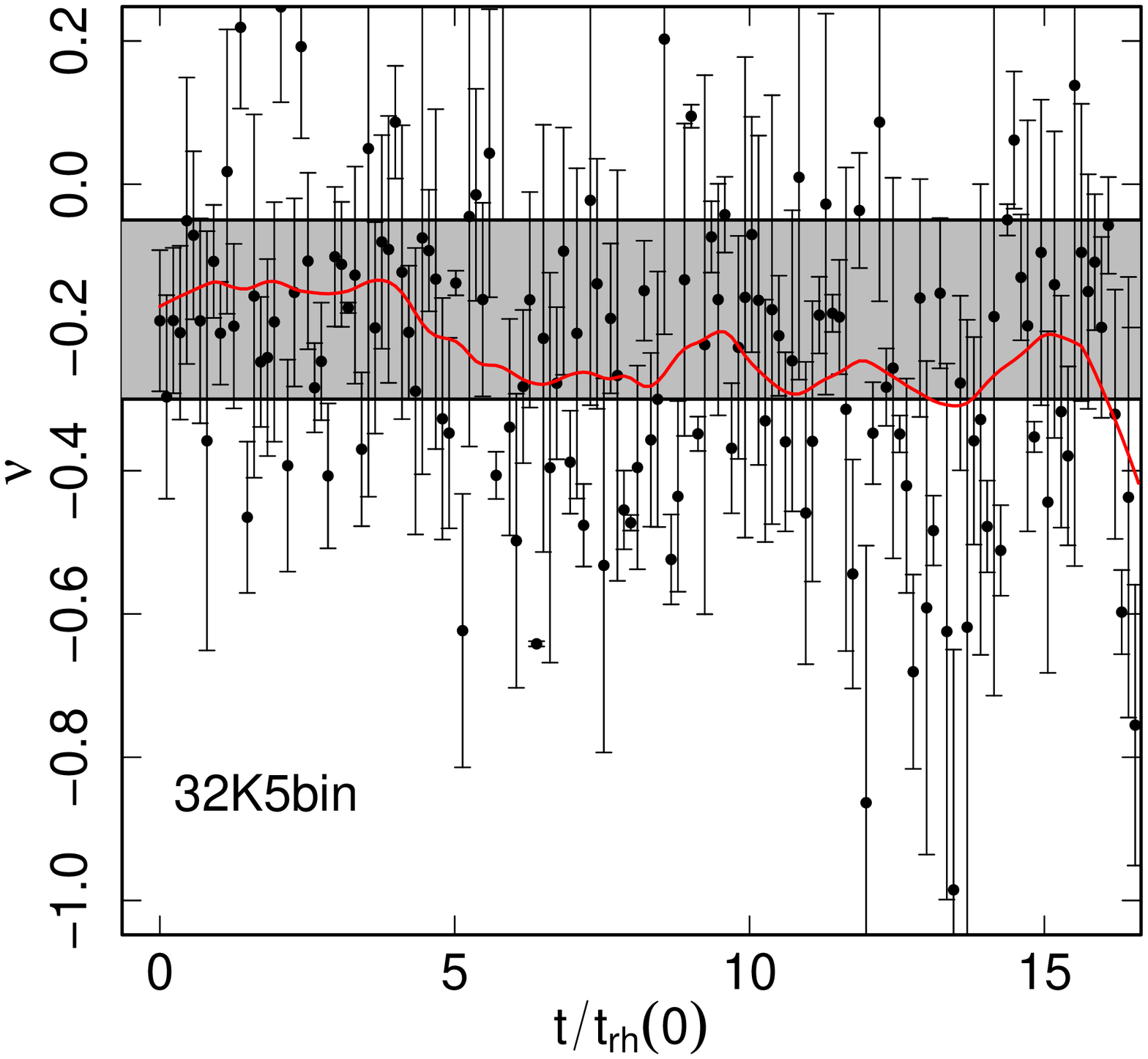}
\includegraphics[width=2.8in]{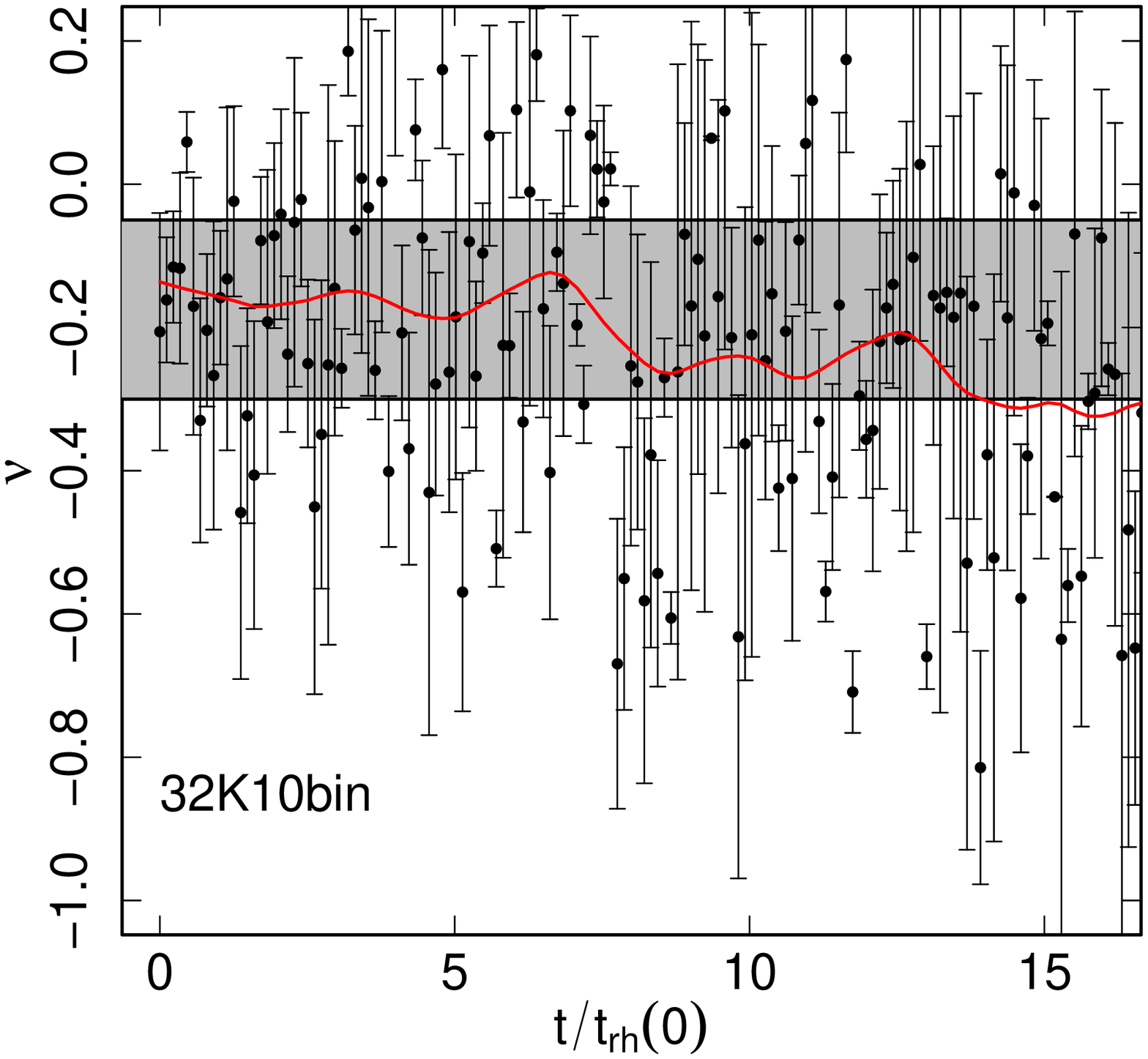}
\caption{Time evolution of the slope, $\nu$, of the inner surface
  brightness profile for the 32KBH (top left), 32K (top right), 32K1bin (middle left) and 32K3bin (middle
  right), 32K5bin (lower left), 32K10bin (lower
  right) simulations. The shaded area shows the  
  range of values of $\nu$ expected for systems harboring an IMBH. The
  solid line in each panel show a smooth fit to the data. 
}  \label{fig:32k}

\end{center}
\end{figure}

The six panels of Fig.~\ref{fig:32k} show the time evolution of the
index $\nu$ for the
simulations with 32K particles. A smooth curve fit to the data has
been added to highlight the evolutionary trend. The grey shaded area
shows the range of values of $\nu$ expected for systems hosting IMBHs
based on past studies (Baumgardt et al. 2005, Miocchi 2007). Our IMBH
run falls within this range presenting a shallow inner cusp with slope
$\nu \sim -0.25$, approximately constant during the entire cluster
evolution.

The results of the simulation 32K (a system with no
IMBH and no primordial binaries), on the other hand, show an evolution
of  $\nu$  
characterized by three 
phases. For $t/t_{rh}(0)\ltorder 5$, the system is
in the pre-core-collapse phase; as anticipated by our preliminary
estimates of $\nu$ for a system described by a King model in
\S\ref{sec:initc}, during this phase the value of $\nu$  undergoes   
significant fluctuations  and, most of the time,  falls 
in the range of values expected for systems hosting an IMBH (either directly or
within 1-sigma).

 The subsequent phase, $ 5 \ltorder
t/t_{rh}(0)\ltorder 7$, is the core collapse phase and is
characterized by a transition to steeper slopes typical of the
post-core-collapse phase. During a significant portion of this
phase the system is characterized by a central shallow cusp with
values of $\nu$ in the range expected for systems with an IMBH.
  Finally for $t/t_{rh}(0)\gtorder 7$ the  system\footnote{Note that
    the values of $t/t_{rh}(0)$ that separates these three phases
    might differ for systems starting with a different initial concentration.}
 is in the post-core-collapse phase and the slope of the inner
  cusp oscillates in the range $\sim -0.4 - -0.7$; these values are
  steeper than those predicted for clusters hosting an IMBH but even
  during this phase, as a consequence of fluctuations and measurement
  errors, $\nu$ may occasionally fall in the range of values expected
  for globular 
  clusters hosting an IMBH.

For systems with primordial binaries, the middle and lower panels of
Fig. ~\ref{fig:32k} show that,  
for a fraction of primordial binaries $f_{bin}\gtorder 0.03$, the
typical values of $\nu$  overlap with the range of values of a system with an
IMBH also for $t/t_{rh}(0)\gtorder 7$ when the core is supported by
the primordial binaries' burning. This makes it impossible to
distinguish whether the central shallow 
cusp is a consequence of the presence of an IMBH or of the presence of
primordial binaries in the cluster core. We point out that the 
small  $f_{bin}$ adopted here is consistent with that
determined observationally for a number of Galactic clusters (see e.g. Davis et
al. 2008).   

\begin{figure}
\includegraphics[width=2.8in]{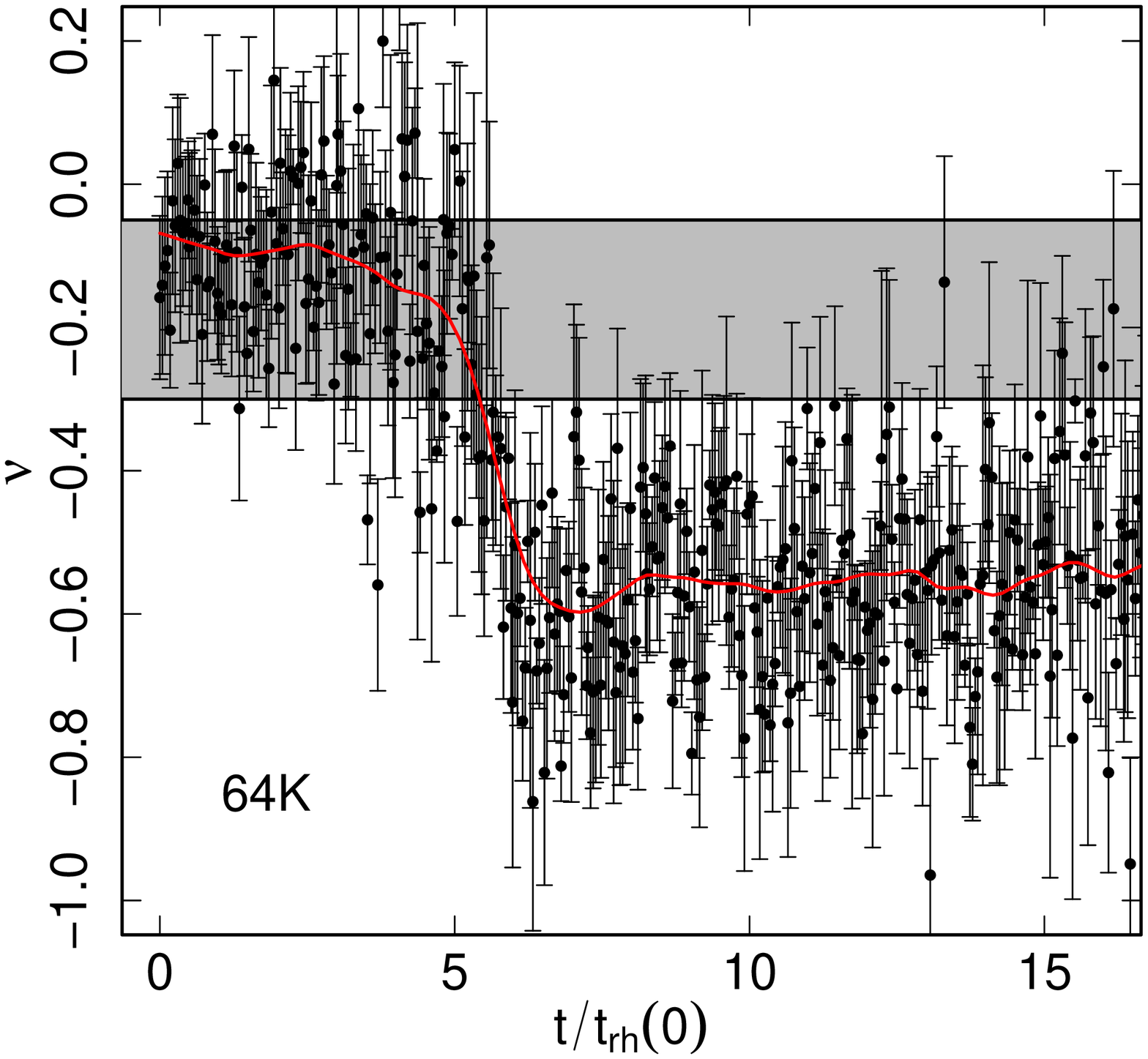}
\includegraphics[width=2.8in]{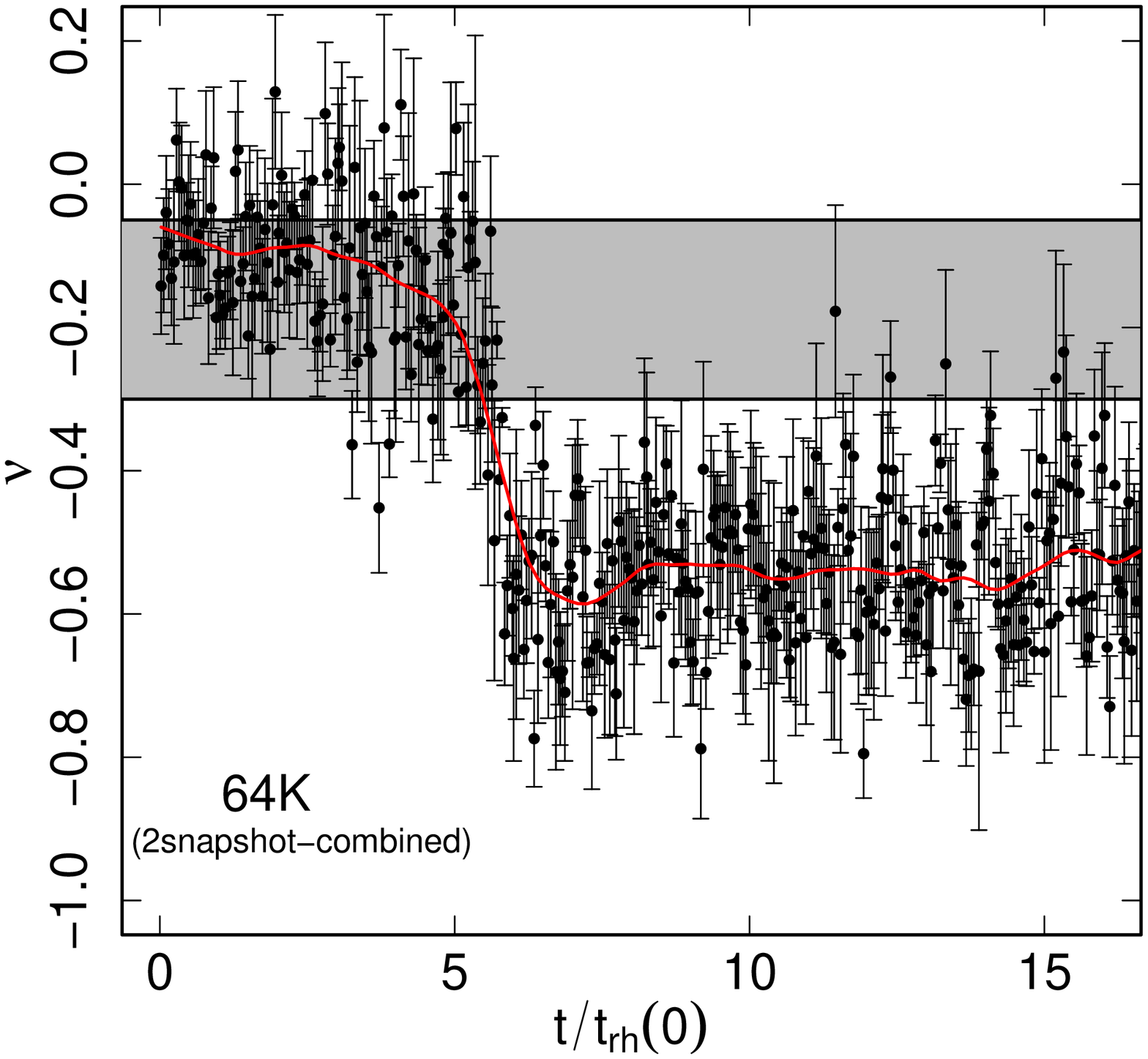}
\includegraphics[width=2.8in]{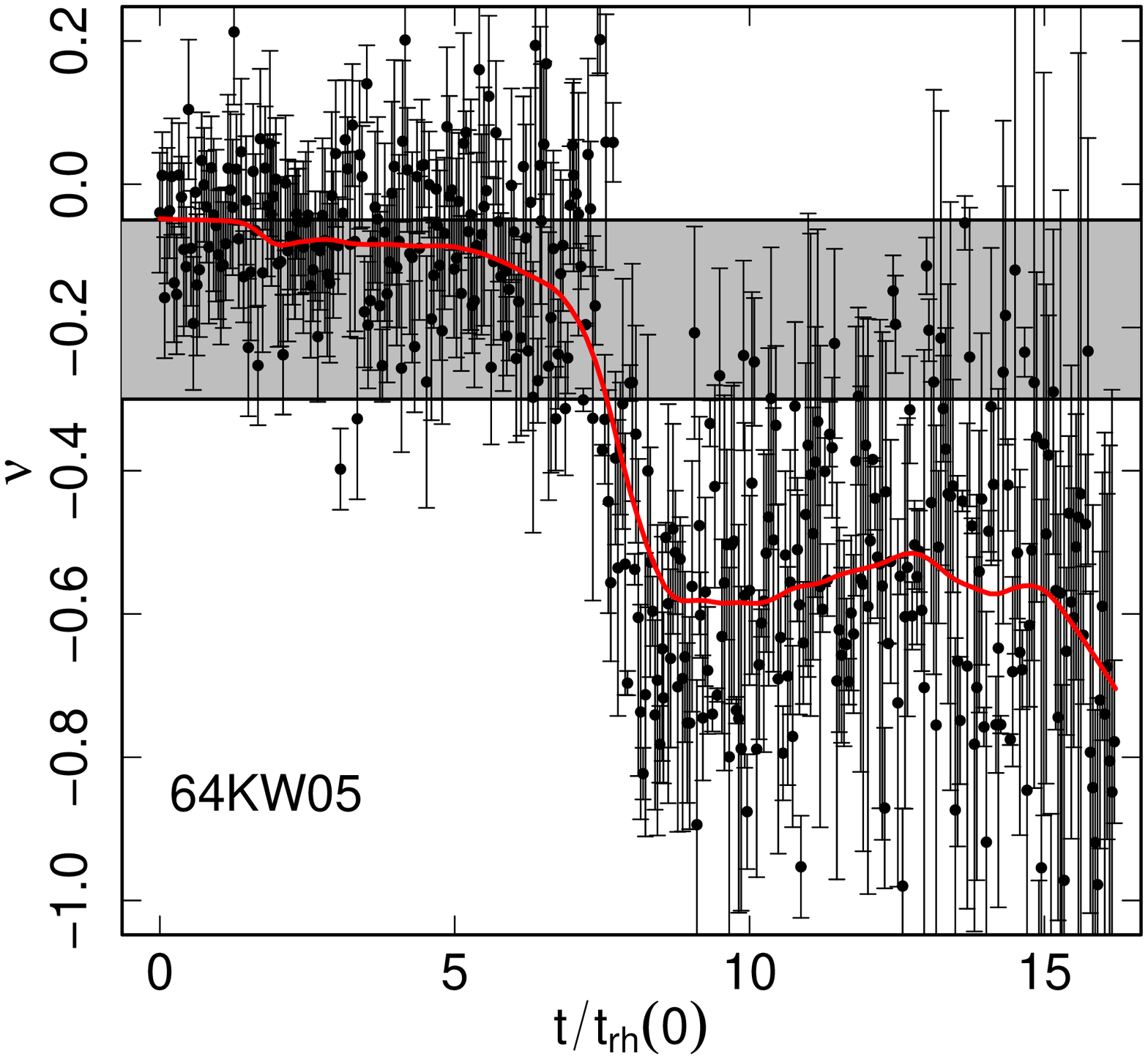}
\includegraphics[width=2.8in]{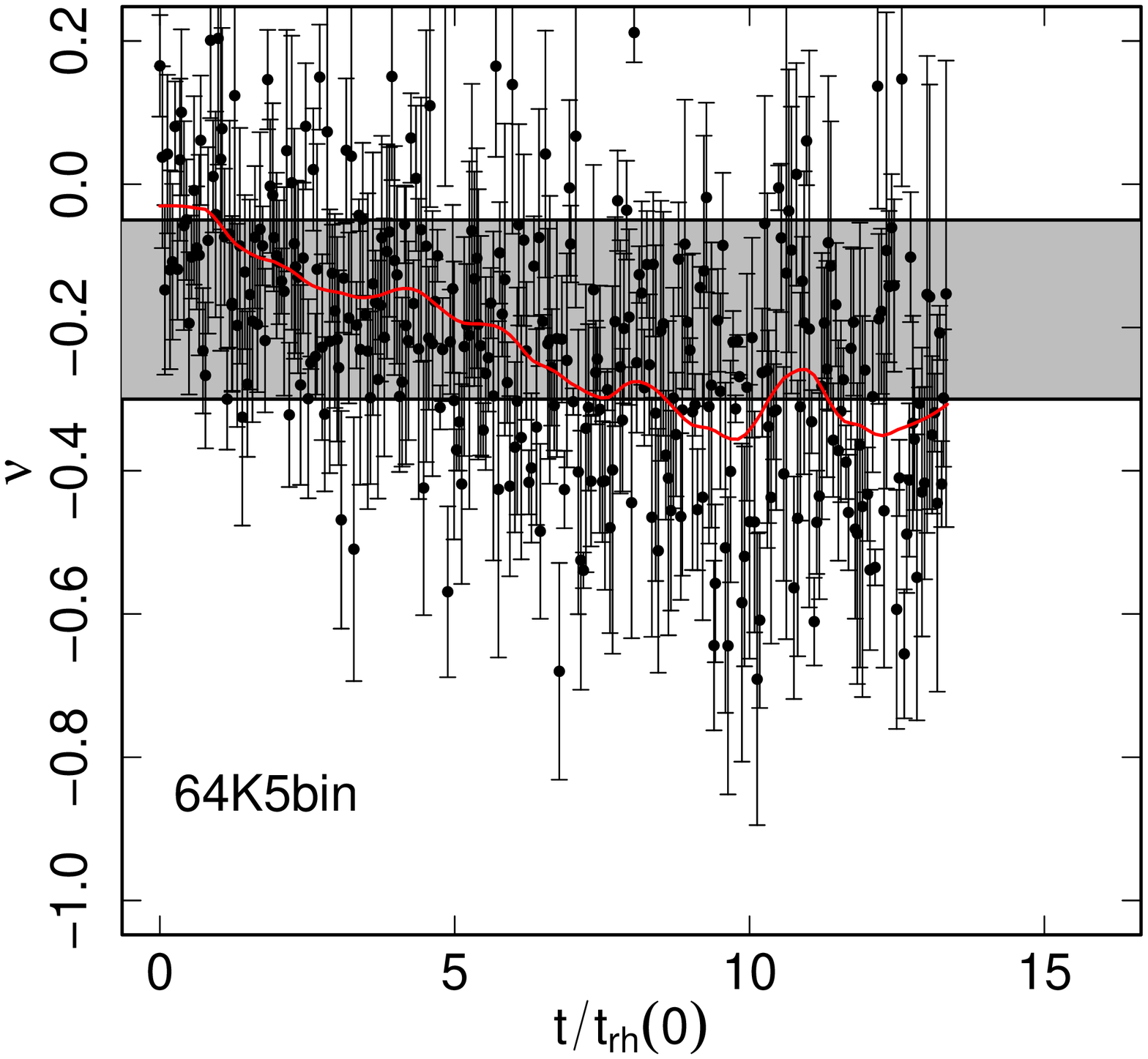}
\includegraphics[width=2.8in]{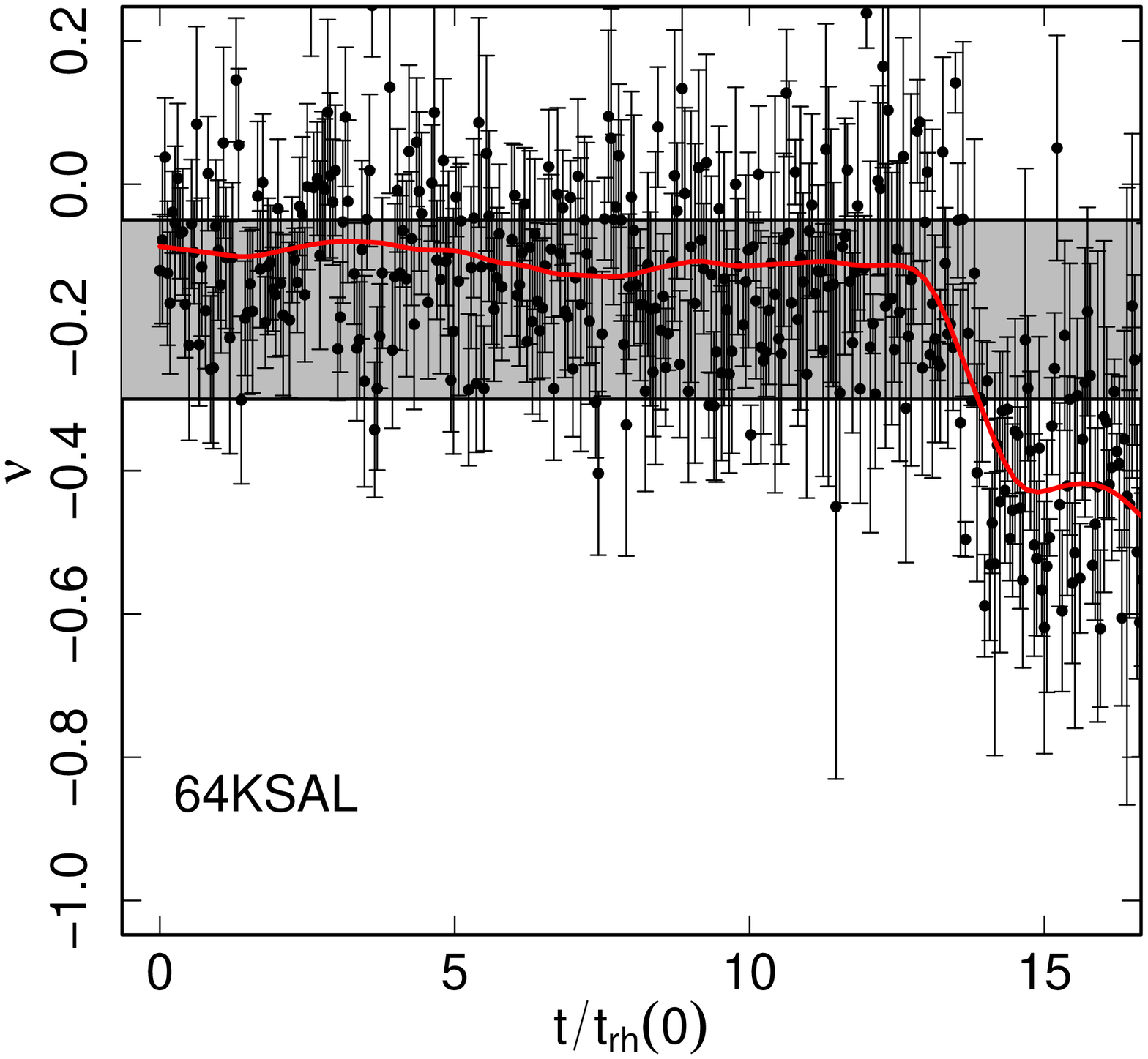}
\caption{Shallow cusp slope $\nu$ versus time as in
  Figure~\ref{fig:32k} for runs: 64K (top left), 64K
  combining two consecutive
  snapshots (top right), 64KW05 (middle left), 64K5bin (middle right) and 64KSAL (lower left).} \label{fig:64k}

\end{figure}

In Figure ~\ref{fig:64k} we show the results of our simulations with
64K particles. Simulation 64K  confirms the results of
the lower resolution run (32K), although the larger number of
particles leads to a smaller average error on the estimate of $\nu$.
In order to further illustrate the reduction in the error on $\nu$ for
an increasing number of particles, we also show  
in Figure ~\ref{fig:64k} a panel with the time evolution of $\nu$
obtained by combining two consecutive snapshots of the simulation
64K. The three phases in
the evolution of $\nu$ described above are evident in both panels.
Other than the smaller error on $\nu$ for larger N, our simulations do not show any dependence of the evolution of $\nu$ on N: for the 32K
run the average $\nu$ is 0.08 with a $1\sigma$ spread equal to 0.17 for
$t/t_{rh}(0)<5$ while the average $\nu$ is 0.52 with a $1\sigma$
spread equal to 0.19 in the post-core-collapse phase $(7<t/t_{rh}(0)<16)$. For the 64K run the average $\nu$ ($1\sigma$ spread) is 0.11 (0.13) for  $t/t_{rh}(0)<5$, and 0.55 (0.13) for $7<t/t_{rh}(0)<16$.

For simulation 64K, the difference between the
measured value of $\nu$ and $\nu=-0.3$ (the lower limit of the range
associated with the presence of an IMBH) is, in the 
post-core-collapse phase, smaller than $1\sigma$  20\% of the
time.
The smaller error on $\nu$ when consecutive snapshots are combined leads to a 
reduced fraction of  post-core-collapse snapshots (10\% of the time
in this case) during which $\nu$ is within $1\sigma$ from   
the black hole range.

Simulation 64KW05 shows a similar time evolution of $\nu$ and the only
difference is the delay in the onset of the core collapse phase,
because of the lower initial concentration.

Simulation 64K5bin also confirms the results of the analogous lower
resolution run (32K5bin).

Finally, the results of the simulation with a Salpeter initial mass
function show that the amount of dark remnants in a cluster plays a
key role in driving the cluster evolution and in determining the shape
of its inner surface brightness profile. In particular our results
confirm that the presence of dark remnants can significantly delay the
onset of core collapse (see e.g. Merritt et al. 2004, Mackey et al.
2008, Trenti et al. 2010) and, in doing so, they extend the
pre-core-collapse phase characterized by small values of $\nu$
consistent with those of systems hosting an IMBH. Dark remnants also play an
important role in the post-core-collapse phase: specifically, by
segregating in the cluster inner regions and heating the visible
cluster population, dark remnants tend to produce a shallower
core-collapse cusp
than that observed in systems with no primordial binaries, no IMBH and
a smaller population of dark remnants.

\section{Discussion}
\label{sec:disc}
Our simulations show that during their evolution,
clusters without an IMBH are often
characterized by central shallow cusps in the surface brightness
profile with a slope falling in the range of values expected for
systems with an IMBH.   

For clusters in  the post-core-collapse phase with neither primordial
binaries nor a significant fraction of dark remnants, the inner cusp
is steeper ($-0.7\ltorder\nu\ltorder-0.4$)  
than that of systems with an IMBH. If we know that a cluster is in the
post-core-collapse phase and has neither
primordial binaries nor a significant fraction of dark remnants, then
a shallow cusp in the range $-0.3\ltorder\nu\ltorder -0.05$  could be
indicative of an IMBH. 

However, a small
fraction of primordial binaries ($f_{bin}\gtorder 0.03$) and/or the
amount of dark remnants produced assuming a Salpeter 
stellar initial mass function (with a 100 per cent retention of dark
remnants) act as a central heating mechanism  
sufficient to produce a shallow inner cusp with the same range of slope
values expected for systems hosting an IMBH.

\begin{figure}
\begin{center}
 \includegraphics[width=6.in]{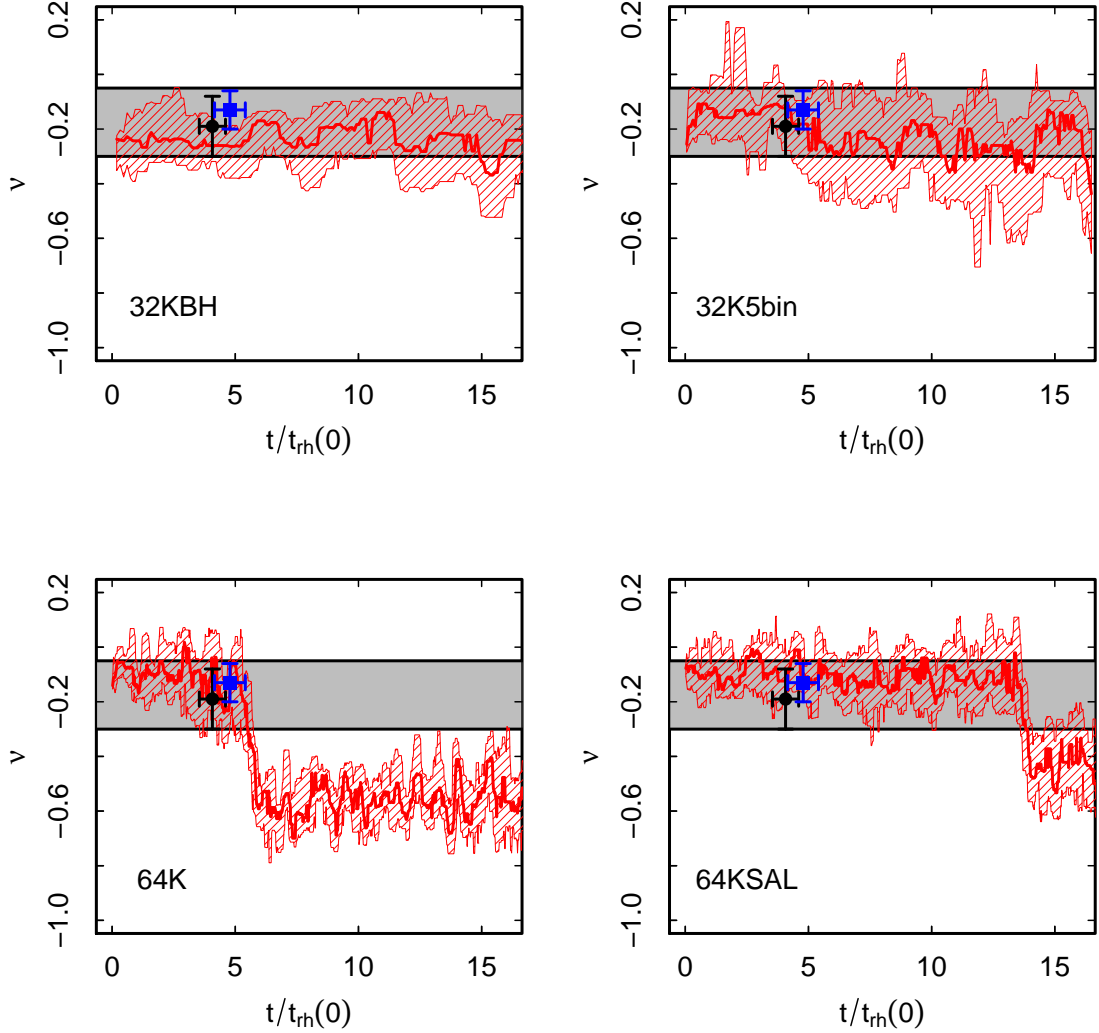}
\caption{$1\sigma$ area of $\nu$ for the 32KBH (top left),
32K5bin (top right), 64K (bottom left), 64KSAL (bottom
right). The upper (lower) line delimiting each   
shaded area represents the 84.1 (15.9) percentile of a moving bin
including 15 values of $\nu$.     
The  dots show the slope of the inner cusp for NGC 6388 (filled
blue square) and NGC 5694 (filled black dot) (values from Noyola \&
Gebhardt, 2006). The dynamical  
ages 
of these two clusters have been calculated assuming $t=11.5 ~\hbox{Gyr} \pm 1.5
~\hbox{Gyr}$ and $t_{rh}(0)$ has been calculated as
  $2t_{rh}$ where $t_{rh}$ is the cluster current half-mass relaxation
time (see section \ref{sec:disc} for further discussion).
  The grey shaded area shows the
  range of values of $\nu$ expected for systems harboring an IMBH.}
\label{fig:summary}

\end{center}
\end{figure}

Figure ~\ref{fig:summary} summarizes the results of a representative
subset of our simulations showing the different dynamical phases and
cluster properties that can be associated with the presence of a
shallow cusp. This figure nicely illustrates the problem faced by any
attempt to use a shallow cusp as an indicator of the presence of an
IMBH.

As an example, we plotted in Fig.~\ref{fig:summary} the Galactic
clusters NGC 5694 and NGC 6388 for which Noyola \& Gebhardt (2006)
estimated, respectively, $\nu=-0.19 \pm 0.11$ and $\nu=-0.13 \pm
0.07$ . Following Hurley (2007), we estimated the dynamical ages of
NGC 5694 and NGC 6388 assuming $t_{rh}(0) \sim 2 t_{rh}(now)$, where
$t_{rh}(now)$ is the current value of the half-mass relaxation
time (from the most recent version of the Harris(1996) catalogue)\footnote{We emphasize  that this
  dynamical age is very uncertain. Specific models for individual
  clusters (see e.g. Giersz \& Heggie 2009) would be necessary for a
  higher accuracy determination of the cluster dynamical age.}. The
values for the slope of the inner cusp of these two 
clusters fall within the range of values expected for systems hosting
an IMBH and indeed Umbreit et al. (2009) reproduced
the NGC5694 surface brightness profile  with a Monte
Carlo simulation of a system hosting an IMBH with a mass $\mbh~=~0.003~
M_{cluster}$. As for NGC 6388, Lanzoni et al. (2007) fit the observed
surface brightness profile with a multimass King model with an IMBH and
suggested the presence of an IMBH with $\mbh \sim 5.7 \times 10^3
M_{\odot}$ ($\sim~0.0022~M_{cluster}$).

In Fig.~\ref{fig:summary} we show, however, that the slopes of the
observed surface brightness profile of these clusters are also
consistent with the results of simulations of systems without an IMBH
under a variety of different initial conditions.
In particular with the estimated dynamical age, these clusters might
be either in the pre-core-collapse phase or currently entering the
core-collapse phase. It is interesting to note that in a recent study
based on radio observations of NGC 6388, Cseh et al. (2010) obtained
an upper limit $\mbh<735 \pm 244 M_{\odot}$, significantly smaller
than the estimate of $\mbh$ based on the surface density profile.

Our analysis clearly shows that shallow cusps in the surface
brightness profile appear naturally during the dynamic evolution of
star clusters without an IMBH. Therefore their presence in an
observed profile cannot be used as a strong argument in favor of the
presence of a central IMBH. 

\acknowledgements

 We
acknowledge support from grants NASA-NNX08AH29G, NASA-NNX08AH15G,
HST-AR-11284, TG-AST090094 and TG-AST090045.

{}
\clearpage

\clearpage

\clearpage

\clearpage

\begin{table}
\begin{center}
\caption{Summary of N-body simulations\label{tab:sim}}
\begin{tabular}{cccccc}
\tableline\tableline
Id. & $N$ & $f_{bin}$ & IMF & $W_0$ & $M_{BH}/M_{cluster}$ \\
32KBH & 32769 & 0.00& MS & 7 & 0.01\\
32K & 32768 & 0.00& MS & 7 &0.0\\
32K1bin & 32768 & 0.01& MS & 7 &0.0\\
32K3bin &32768 & 0.03& MS & 7 &0.0\\
32K5bin & 32768 & 0.05& MS & 7 &0.0\\
32K10bin & 32768 & 0.10& MS & 7 &0.0\\

64KW05& 65536 & 0.00& MS & 5 &0.0\\
64K&65536 & 0.00& MS & 7 &0.0\\
64K5bin&65536 & 0.05& MS & 7 &0.0\\
64KSAL&65536 & 0.00& Sa & 7 &0.0\\
\tableline
\end{tabular}
\tablecomments{Summary of initial conditions of the N-body simulations presented in this Letter. First column: Identification; second column:
  number of particles $N$; third column:
  binary fraction $f_{bin}=N_b/N$; fourth column: initial mass function used
  (Sa: Salpeter, MS: Miller \& Scalo); fifth column: initial
  concentration of the density profile (King index $W_0$); sixth column: IMBH mass in units of the total cluster initial mass.}
\end{center}
\end{table}

\end{document}